# Ventilator pressure prediction using recurrent neural network

Su Diao,
Department of Industrial & Systems Engineering, Auburn University,
Alabama, USA, 36849
scottdiao33@gmail.com

Changsong Wei,
Digital Financial Information Technology Co.LTD
Sichuan, China, 610041
changsongwei88@gmail.com

Junyu Wang
Computer Science Department, University of California,
Los Angeles, California, USA, 90095,
junyu08@cs.ucla.edu

Yizhou Li
Department of Electrical, Computer, and Systems Engineering, Case Western Reserve University
Cleveland, USA, 44106
yxl3527@case.edu

*Abstract*—This paper presents a recurrent neural network approach to simulating mechanical ventilator pressure. The traditional mechanical ventilator has a control pressure that is monitored by a medical practitioner and can behave incorrectly if the proper pressure is not applied. This paper takes advantage of recent research and develops a simulator based on a deep sequence model to predict airway pressure in the respiratory circuit during the inspiratory phase of a breath given a time series of control parameters and lung attributes. This method demonstrates the effectiveness of neural network-based controllers in tracking pressure wave forms significantly better than the current industry standard and provides insights into the development of effective and robust pressure-controlled mechanical ventilators. The paper will measure as the mean absolute error between the predicted and actual pressures during the inspiratory phase of each breath.

*Keywords—Ventilator pressure, recurrent neural network, mechanical ventilator pressure, control parameters, lung attribute*

## I. Introduction (Heading 1)

What does a doctor do when a patient is having difficulty breathing? They use a ventilator to pump oxygen into a patient's lungs through a tube in the patient's windpipe while the patient is sedated. Mechanical ventilation, on the other hand, is a time-consuming procedure that requires the involvement of a clinician, a limitation that was prominently displayed during the early days of the COVID-19 pandemic. Developing new methods for controlling mechanical ventilators, on the other hand, is prohibitively expensive, even before reaching the stage of clinical trials. High-quality simulators have the potential to lower this barrier.

To train current simulators, they must be used in an ensemble setting, with each model simulating a single lung setting. However, because lungs and their attributes exist in a continuous space, it is necessary to investigate a parametric approach that takes into account the differences in patient lungs.

By collaborating with Princeton University, the Google Brain team hopes to expand the community of people who are interested in machine learning applications for mechanical ventilation control. Rather than the current industry standard of PID controllers, they believe that neural networks and deep learning will be able to generalise across lungs with varying characteristics more effectively.

In this paper, we will simulate a ventilator connected to a patient's lung while the patient is under anaesthesia. We will also take into consideration the lung characteristics of compliance and resistance. It will aid in the reduction of the cost barrier associated with the development of new methods of controlling mechanical ventilators. In the future, algorithms that adapt to patients will be developed, which will relieve clinicians of some of the burden they are currently under during these novel times and beyond. As a result, it is possible that more patients will have access to ventilator treatments to assist them in breathing.

The mechanical ventilator is an essential medical component in an intensive care setting, particularly during the COVID-19 pandemic. The current industry standard for PID controllers employs a combination of proportional, integral, and derivative controls that use the deviation from the target waveform as input and adjust it in the pressure chamber to correct the output waveform and close the deviation gap. As a result, the PID controller relies on continuous manual monitoring and adjustment by medical practitioners, resulting in incorrect pressure. While it meets the needs and safety requirements, it is unable to generalise and adapt quickly to different clinical conditions.

As a result, a dynamic controller that constantly adjusts its pressure may be able to solve this problem. That is where machine learning enters the picture. Machine learning models, as is well known, are data-hungry and require a large amount of data that is difficult to obtain. As a result, two types of machine learning models are required in this case: the first model is a simulator that generates data, allowing the second model to be fine-tuned using this data and replacing the traditional PID controller algorithm. A recent paper (Suo et al., 2021) demonstrated the potential benefit of machine learning for ventilator control on an open-source ventilator (LaChance et al., 2020) designed in response to the COVID-19 pandemic, paving the way for intelligent control methods that are robust and require less manual monitoring. We take advantage of this research by presenting a simulator based on a deep sequence model and utilising statistical learning techniques to improve the model's enhancement.

Even though it appears unlikely that AI will ever completely replace professional health care workers, it is advantageous to use computing power to analyse "big data" for the benefit of patients.

Deep Learning (DL) methods based on recurrent neural networks are now commonly used to solve complex mathematical problems, particularly those with temporal dependencies. RNN weighted input values are summarised and repeatedly updated to produce an output that best reflects the outcome of interest. Furthermore, recurrent feedback mechanisms produce a memory function. Hoch Reiter and Schmidhuber's Long Short-Term Memory model (LSTM) solves complex tasks by using a constant error fow ("constant error carousels") within memory cells with an opening and closing gate function, enabling quasi-sustained short-term memory. Since their introduction, RNNs, particularly LSTMs, have been used for a variety of tasks such as handwriting recognition and speech recognition, as well as in a variety of healthcare applications. Machine Learning (ML) already has a greater impact on daily life than we may realise, and it is critical for the technology industry.

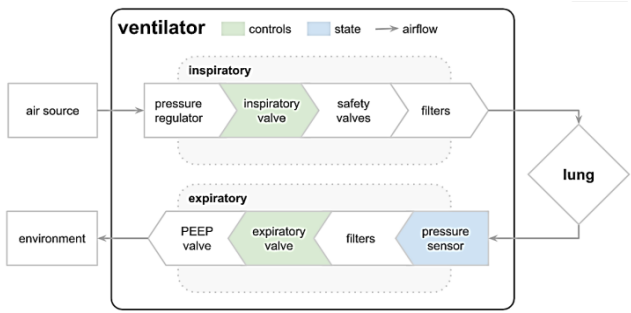

Fig. 1.  Ventilator Blueprint

Under certain conditions, the concept of ML for analysing complex and frequently highly heterogeneous patient collectives appears reasonable in critical care medicine. Several studies have evaluated the use of ML for the treatment of sepsis, assessing patient prognosis and/or risk for prolonged clinical courses, and a variety of other applications.

Various studies have been conducted that demonstrated that ML can be used as a prognostication tool for ICU-mortality and/or assessment of patients on mechanical ventilation by AI. In their study on 20,262 ICU stays not included in the MIMIC-III database, Parreco et al. were able to reliably identify patients at risk for tracheostomy and prolonged MV. Chen et al. used ML to analyse sensor arrays on exhaled breath samples to detect ventilator-associated pneumonia in patients on MV. Several other studies with promising results have been conducted in this field, making the use of ML in clinical daily routine on the ICU likely in the future.

A. INSTRUMENTS USED
  1) PID CONTROLLER
A PID controller is a type of instrument that is used in industrial control applications to regulate variables such as temperature, flow, pressure, speed, and other parameters. PIDC controllers, which stand for proportional integral derivative, control process variables through a feedback mechanism known as a control loop feedback mechanism. They are the most accurate and stable of the controllers available. The operation of a PID is explained in greater detail in this article.

PID control is a well-established method of guiding a system towards a desired position or level by adjusting its parameters. It is practically ubiquitous as a means of controlling temperature, and it finds application in a wide range of chemical and scientific processes, as well as in automated systems and robotics systems. Precision inertial control (PID) is a closed-loop control technique that attempts to maintain an output from a process as close as possible to the target or setpoint output.

When a pandemic strikes, the commercial sector has safe and reliable ventilation technology; however, the small number of capable suppliers is unable to keep up with the high demand for ventilators during the outbreak. Apart from that, expensive and inaccessible specialised, proprietary equipment developed by medical device manufacturers is prohibitively expensive and unavailable in low-resource settings. The People's Ventilator Project (PVP) is an open-source, low-cost pressure-control ventilator that is designed to be as independent of specialised medical parts as possible, allowing it to better adapt to supply chain shortages in the event of a disaster. Generally speaking, the PVP adheres to established design conventions, with the most notable exception being active and computer-controlled inhalation, which is combined with passive exhalation. It supports pressure-controlled ventilation in conjunction with standard features such as autonomous breath detection and a comprehensive set of FDA-mandated warnings and alarms.

Hardware PVP is a pressure-controlled ventilator that makes use of a small number of low-cost, off-the-shelf hardware components to achieve its function. The inspiratory flow is controlled by a low-cost proportional valve, and the expiratory flow is controlled by a relay valve. In addition to measuring airway pressure with a gauge pressure sensor, an inexpensive D-lite spirometer used in conjunction with a differential pressure sensor measures expiratory flow with a differential pressure sensor.

Pi 4 boards, which run the graphical user interface, administers the alarm system, monitors sensor values, and sends actuation commands to the valves. The Raspberry Pi 4 board is in charge of coordination of the PVP's components, which are all coordinated by a single Raspberry Pi 4 board. The Raspberry Pi's electrical system is comprised of two modular board 'hats,' a sensor board and an actuator board, which are connected together by 40-pin stackable headers to form the core of the system. Because of the modularity of this system, individual boards can be revised or modified to substitute components in the event of a shortage of a particular component.

B. SOFTWARE

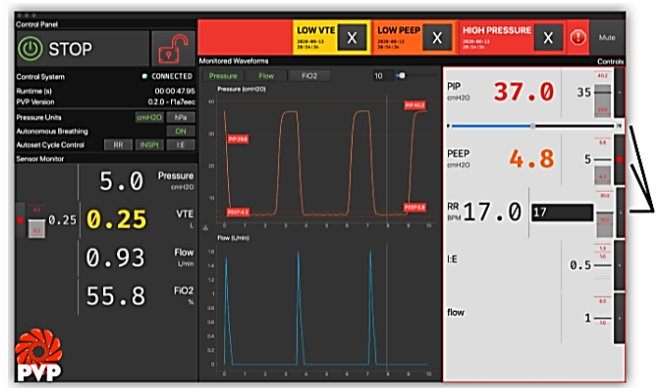

Fig. 2.

The software developed by VP was created with the goal of bringing the philosophy of free and open-source software to medical devices. The PVP is not only completely open from top to bottom, but we have also designed it to serve as the framework for an adaptable, general-purpose ventilator that has been developed collaboratively. Ventilation control system written entirely in high-level Python (3.7) eliminates the development and inspection bottlenecks associated with split computer/microprocessor systems, which require users to read and write low-level hardware firmware in order to function.

All of PVP's components are modular in design, allowing them to be reconfigured and expanded to accommodate new ventilation modes and hardware configurations as they become available. Our complete API-level documentation, along with an automated testing suite, allows anyone to inspect, understand, and extend the functionality of PVP's software framework.

### C. QUICK LUNG

Easy to use, the Quick Lung is a precision test lung that can be adjusted for different levels of difficulty. Because of its small size and ease of use, it is capable of simulating a wide range of patient conditions, including patient inspiratory efforts, in a single device. Quick Lung is available in two models: an adult model and a paediatric model, both of which are calibrated and standard. More information can be found in the "Options" section below. You can use Quick Lung on its own, or you can customise it to meet your specific requirements by adding features such as spontaneous breathing. All ventilators, from ICU to transport, are compatible with this product.

*1) The Purpose of the Quick Lung*
- Integrate any flow/volume/pressure analyser to create a complete ventilator testing system for verification of ventilator performance.
- Instructional sessions in-service and a sales demonstration
- Mechanical ventilation training at the fundamental level

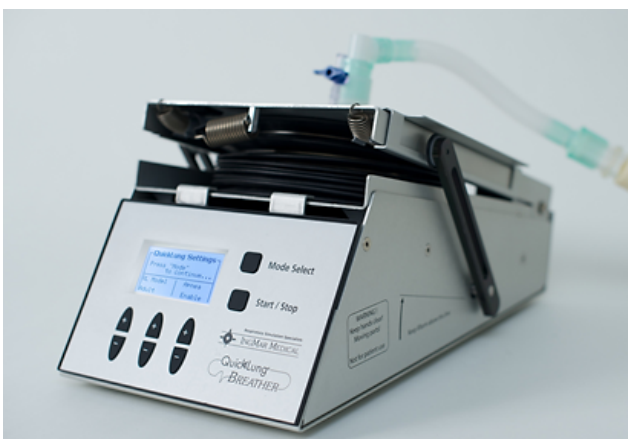

Fig. 3. Quick Lung

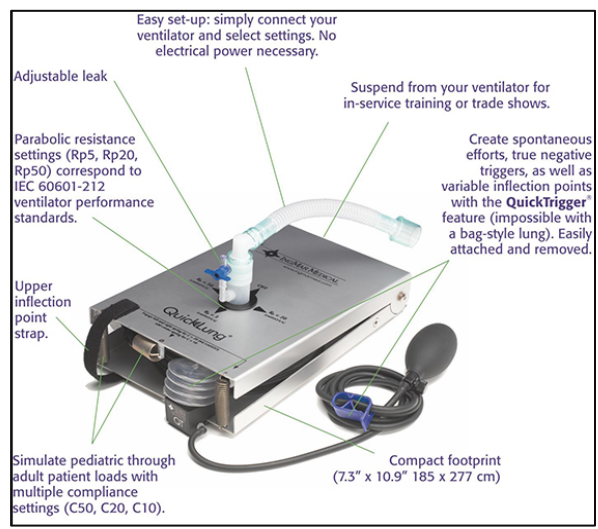

Fig. 4. Quick Lung working mechanism

## II. DATASET DESCRIPTION

Quick look at the training data

TABLE I.

| Id | Breath_id | R | C | Time_step | U_in | U_out | Pressure |
|---|---|---|---|---|---|---|---|
| 1 | 1 | 20 | 50 | 0 | 0.083334 | 0 | 5.837492 |
| 2 | 1 | 20 | 50 | 0.033652 | 18.38304 | 0 | 5.907794 |
| 3 | 1 | 20 | 50 | 0.067514 | 22.50928 | 0 | 7.876254 |
| 4 | 1 | 20 | 50 | 0.011542 | 22.80882 | 0 | 11.74287 |
| 5 | 1 | 20 | 50 | 0.135756 | 25.35585 | 0 | 12.23499 |
| ... | ... | ... | ... | ... | ... | ... | ... |
| 6035996 | 125749 | 50 | 10 | 2.504603 | 1.489714 | 1 | 3.869032 |
| 6035997 | 125749 | 50 | 10 | 2.537961 | 1.488497 | 1 | 3.869032 |
| 6035998 | 125749 | 50 | 10 | 2.571408 | 1.558978 | 1 | 3.798729 |
| 6035999 | 125749 | 50 | 10 | 2.604744 | 1.272663 | 1 | 4.079938 |
| 603600 | 125749 | 50 | 10 | 2.638017 | 1.482739 | 1 | 3.869032 |

### A. COLUMNS

Id represents globally-unique time step identifier across an entire file. Breath_id here is globally-unique time step for breaths R is for lung attribute indicating how restricted the airway is (in cmH2O/L/S). Physically, this is the change in pressure per change in flow (air volume per time). Intuitively, one can imagine blowing up a balloon through a straw. We can change R by changing the diameter of the straw, with higher R being harder to blow. C is for lung attribute indicating how compliant the lung is (in mL/cmH2O). Physically, this is the change in volume per change in pressure. Intuitively, one can imagine the same balloon example. We can change C by changing the thickness of the balloon's latex, with higher C having thinner latex and easier to blow.

Timestamp shows the actual time stamp. u_in is the control input for the inspiratory solenoid valve. Ranges from 0 to 100. U_out is for the control input for the exploratory solenoid valve. Either 0 or 1. Pressure here is for the airway pressure measured in the respiratory circuit, measured in cmH2O. Data. Shape is for unique values do we have for each feature.

## B. DATASET DESCRIPTION OF TRAINING DATASET

TABLE II.

|  | 0 |
|---|---|
| Breath _id | 75450 |
| R | 3 |
| C | 3 |
| Time_step | 3767571 |
| U_in | 4020300 |
| U_out | 2 |
| Pressure | 950 |

Dataset Description of Testing Dataset

TABLE III.

|  | 0 |
|---|---|
| Breath _id | 50300 |
| R | 3 |
| C | 3 |
| Time_step | 2855528 |
| U_in | 2787822 |
| U_out | 2 |
|  | 0 |

We can see that we have over 6 million rows of training data, corresponding to 75,450 breaths, and 50,300 breaths in the test dataset. On average we have 80 time steps of data per breath. Let us check this for the training data.

In this data the unit of time is seconds. Here we will see how long does the breath lasts 2.9372379779815674 Here we counted that the longest breath is just under 3 seconds. The maximum time that the exploratory solenoid valve is set to 0 and it is 0.999798059463501

The valve seems to be activated after 1 second.

TABLE IV.

| Id | Breath _id | R | C | Time_step | U_in | U_out | Pressure |
|---|---|---|---|---|---|---|---|
| 0 | 1 | 20 | 50 | 0.000000 | 0.083334 | 0 | 5.837492 |
| 1 | 1 | 20 | 50 | 0.033652 | 18.383041 | 0 | 5.907794 |
| 2 | 1 | 20 | 50 | 0.067514 | 22.509278 | 0 | 7.876254 |
| 3 | 1 | 20 | 50 | 0.011542 | 22.808822 | 0 | 11.742872 |
| 4 | 1 | 20 | 50 | 0.135756 | 25.355850 | 0 | 12.234987 |
| … | … | …. | …. | …. | …. | …. | …. |
| 75 | 1 | 20 | 50 | 2.504603 | 1.489714 | 1 | 3.869032 |
| 76 | 1 | 20 | 50 | 2.537961 | 1.488497 | 1 | 3.869032 |
| 78 | 1 | 20 | 50 | 2.571408 | 1.558978 | 1 | 3.798729 |
| 79 | 1 | 20 | 50 | 2.604744 | 1.272663 | 1 | 4.079938 |
| 75 | 1 | 20 | 50 | 2.638017 | 1.482739 | 1 | 3.869032 |

Here we are visualizing `u_in`, `u_out` and `pressure` with respect to the `time_stamp`:

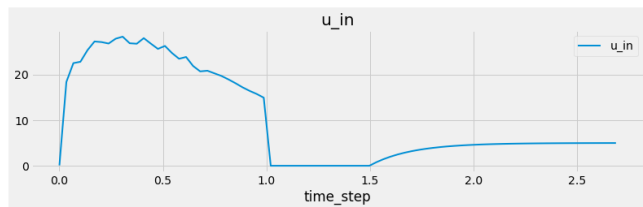

Fig. 5. Visualization of `u_in`

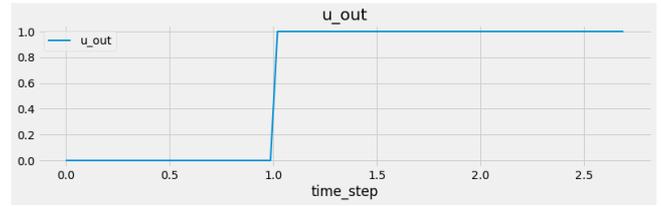

Fig. 6. Visualization of `u_out`

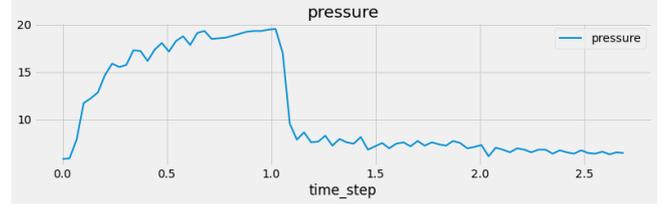

Fig. 7. Visualization of `u_out`

## III. PRESSURE

Here we will look at the `pressure`. The pressure is measured in cmH20, where 1 cmH20 is roughly equal to 98 Pascal's. The global peak inspiratory pressure (PIP) in the training data is 64.8209917386395.

'Pressure' and 'time step'

'Pressure' and 'u_out'

Splitting of categorical and numerical data

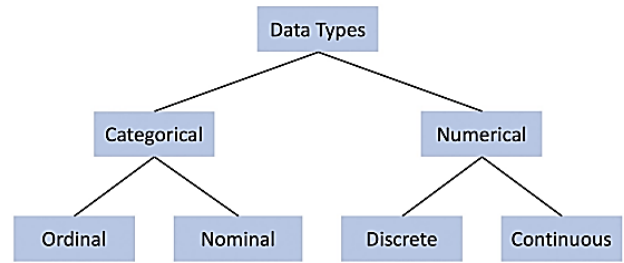

Fig. 8. Splitting of categorical data

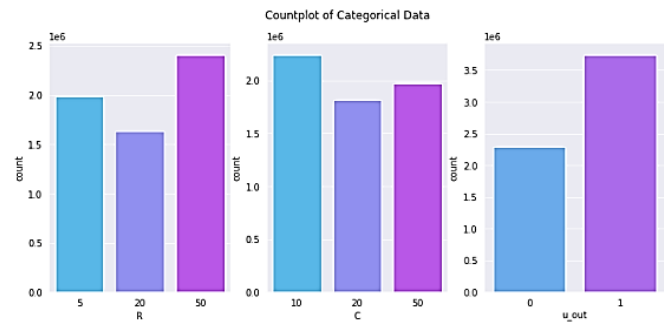

Fig. 9. Categorical data counterplot

## IV. MODEL ARCHITECTURE

### A. ELECTRONIC DESIGN AUTOMATION

In recent years, machine learning for electronic design automation (EDA) has emerged as a popular topic, with numerous studies proposing to use machine learning to improve EDA methods. These studies cover almost all stages of the chip design flow, including design space reduction and exploration, logic synthesis, placement, routing, testing, verification, and manufacturing, among others. When compared to traditional methods, these machine learning-based methods have demonstrated significant improvement.

In order to fully comprehend the data that we are working with, we must first identify any hidden patterns in the data. Exploratory Data Analysis can assist us in determining the correlation between different columns of the data, as well as in analysing the properties of the data. EDA typically consumes approximately 30% of the total project time because we must write a significant amount of code in order to create various types of visualisations and analyse them.

Python provides a large number of libraries that assist in automating the process of EDA, which in turn saves time and effort. However, deciding which library to use can be difficult. The type of problem we are attempting to solve determines which library we should use.

If we are attempting to build a Machine Learning model from the ground up, we can use MLJAR-Supervised to assist us. It is a free and open-source Python library that provides a variety of features, including:

Automating EDA

ML model selection and hyper parameter tuning among others.

Creating reports, and so on.

In this competition we are provided with 75,450 non-contiguous cycles (each cycle is uniquely labelled with an individual breath_id) of the PVP1 automated ventilator connected to a high-grade test lung (Quick lung, Ingmar Medical) Three different values of the compliance (C) were tested mL cm H2O in conjunction with three different values of resistance (R) cm H2O/L/s, resulting in a total of 9 different lung settings.

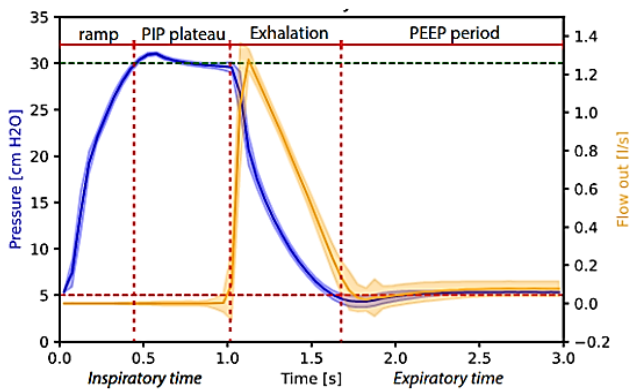

Fig. 10. A typical breath cycle

A cycle lasts for up to 3 seconds. It is the inspiratory section (from 0-1 seconds) that we model in this competition.

When it comes to model evaluation we have to predict the pressure for 50,300 test cycles, of which 19% are assigned to the Public Leader board, and the remaining 81% to the Private Leader board. It is the mean absolute error (mae) between the predicted and actual pressures during the inspiratory phase of each breath that constitutes the evaluation metric in this competition.

As we can see, we have over 6 million rows of training data, which corresponds to 75,450 breaths, and 50,300 breaths in the test dataset, which corresponds to 50,300 breaths. Per breath, we collect an average of 80 time steps of information. Let's see if this holds true for the training data.

## V. ALL BREATHS

What values do we have for R, which represents how restricted the airway is (in cmH2O/L/S).

TABLE V.

|    | R       |
|----|---------|
| 50 | 2410080 |
| 5  | 1988800 |
| 20 | 1637120 |

Now for the values of C, the lungs attribute indicating how compliant the lung is (in mL/cmH2O) *Now for the values of C, the lungs attribute indicating how compliant the lung is (in mL/cmH2O)*

TABLE VI.

|    | C       |
|----|---------|
| 10 | 2244720 |
| 50 | 1971680 |
| 20 | 1819600 |

As a result, we have nine different R and C combinations. To illustrate, let us examine the number of times each of these combinations occurred in the training data (divided by 80 to account for the number of time steps in each breath).

TABLE VII.

| C  | 10     | 20     | 50     |
|----|--------|--------|--------|
| R  |        |        |        |
| 5  | 8312.0 | 8277.0 | 8271.0 |
| 20 | 6070.0 | 6208.0 | 8186.0 |
| 50 | 13677.0| 8260.0 | 8189.0 |

And similarly, for the test data

TABLE VIII.

| C  | 10     | 20     | 50     |
|----|--------|--------|--------|
| R  |        |        |        |
| 5  | 5437.0 | 5451.0 | 5447.0 |
| 20 | 4292.0 | 40.88.0| 5500.0 |
| 50 | 9081.0 | 5503.0 | 5501.0 |

We also have u_out, the control input for the exploratory solenoid valve. Either 0 or 1.

TABLE IX.

|   | U_out   |
|---|---------|
| 1 | 3745032 |
| 0 | 2290968 |

*A. Pressure*

And now we shall look at the pressure. The pressure is measured in cmH20, where 1 cmH20 is roughly equal to 98 Pascal's. The global peak inspiratory pressure (PIP) in the training data is.

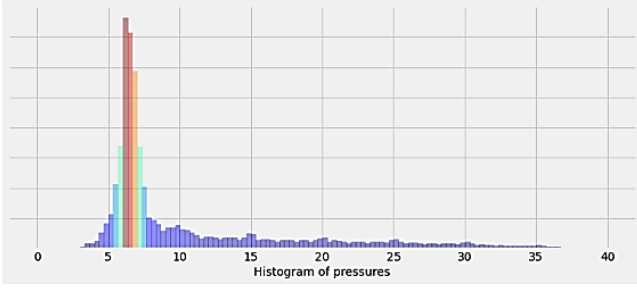

Fig. 11. Histogram of pressure

The pressure recorded at 64.8209917386395

Note however that in this competition the expiratory phase is not scored, so for practical purposes we are only really interested in the pressure for u_out=0, i.e. the first second of the experiments:

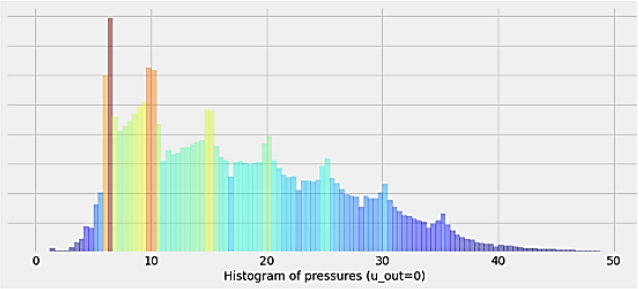

Fig. 12. Pressure histogram

The graph pressure recorded with a median value of 15.82039635914182

## VI. MODEL LSTM

In the following, we will be building on the LSTM with forget gates (Gers et al., 2000), which will be referred to as "LSTM" throughout. The memory block is the fundamental unit of an LSTM network. It contains one or more memory cells as well as three adaptive, multiplicative gating units that are shared by all cells in the network a stumbling block. Each memory cell is composed of a recurrently self-connected linear unit at its heart. The "Constant Error Carousel" is what we call it (CEC). This is accomplished by recirculating activation and error signals. The CEC provides short-term memory storage for extended time periods indefinitely, and it can be used indefinitely. The input, forget, and output gates can all be trained to learn, respectively, what information to accept and what information to reject.

How much data to store in memory, how long to store it for, and when to read it out Memory combining is a process that takes place in a computer? Organizing cells into blocks allows them to share the same gates (provided the task permits this), resulting in faster processing lowering the number of adaptive parameters to be used throughout this paper, j refers to memory blocks, and v refers to memory cells in block j (within the same block). Sj cells), in order that cv j is the v-th cell of the j-th memory block; wlm is the weight on the j-th memory block the connection that exists between units m and l The index m encompasses all of the source units, as specified by the network topology; if the activation of a source unit ym(t1) refers to the activation of an input unit, current

In its place, the external input ym(t) is used. The output yc of memory cell c is calculated on the basis of the input yc the current cell state sc, as well as four different sources of information: The input to the cell itself is represented by zc, and the output is represented by zin. In the input, forget, and output gates, respectively, z and zout provide input and output to the gates. It operates in discrete time steps t = 0, 1, 2,..., with each step involving the updating of all units' activation (forward pass), followed by the computation of error signals for all weights (backward pass) (backward pass).

### A. FORWARD PASS

#### 1) INPUT

During each forward pass we first calculate the net cell input.

$$Z\ C_j^v(t) = \sum_m w\ C_j^v\ m\ y_m(t-1) \quad (1)$$

The input squashing function g is then applied to it as an optional option It is necessary to multiply the result by the activation of the input gate of the memory block, which is calculated by applying a logistic sigmoid squashing function fin with range to the net input zinc of the gate:

$$Y_{inj}(t) = f_{in}(t)), Z_{in}(t) = \sum_m w\ C_j^v\ m\ y_m(t-1) \quad (2)$$

It is the activation yin of the input gate that multiplies the input to all cells in the memory block, and it is this factor that determines which activity patterns are stored (added) into the memory block. During training, the input gate learns to open (yin 1) in order to store relevant inputs in the memory block, and to close (yin 0) in order to shield it from irrelevant ones, respectively. The condition of the cell at time zero, the activation (or state) sc of a memory cell c is initialised to zero; after that, the CEC accumulates a sum, which is discounted by the forget gate, over its input; and finally, the CEC is reset to zero.

To be more specific, we first determine the activation of the forget gate in the memory block.

$$y_{\emptyset j} = (t) = f_{\emptyset j}(z_{\emptyset j}(t) = \sum_m w_{\varphi m} y_m(t-1) \quad (3)$$

where fφ is a logistic sigmoid function with range .The new cell state is then obtained by adding the squashed, gated cell input to the previous state multiplied by the forget gate activation:

$$Sc_j^v(t) = y_{\varphi j}(t) Sc_j^v(t-1) + y_{in}(t) g\left(zc_j^v(t)\right),$$
$$Sc_j^v(0) = 0 \quad (4)$$

As a result, activity continues to circulate in the CEC as long as the forget gate is left open (y 1). A similar process to learning what to store in the memory block occurs when learning about how long to retain the information and, when the information becomes outdated, learning how to erase it by resetting the cell state to zero. Gers and colleagues (2000) discovered that this prevents the cell state from growing indefinitely and that it allows the memory block to store new data without being interfered with by previous operations.

In many ways, LSTMs outperform conventional feed-forward neural networks and recurrent neural networks (RNNs). This is due to their ability to selectively remember patterns over long periods of time, which they have demonstrated. The purpose of this article is to explain LSTM and provide you with the knowledge necessary to apply it to real-world problems.

## VII. RESULTS & CONCLUSION

There are a variety of methods for comparing forecasts with their eventual outcomes, and the mean absolute error is one of them. The mean absolute scaled error (MASE) and the mean squared error (MSE) are two well-established alternatives to consider. All of these measures summarise performance in ways that do not take into consideration the direction of over- or under-prediction; the mean signed difference, on the other hand, does take this into consideration. Whenever a prediction model is to be fitted using a selected performance measure, in the same way that the least squares approach is related to the mean squared error, the equivalent for mean absolute error is the smallest possible number of absolute deviations (also known as least absolute deviations).

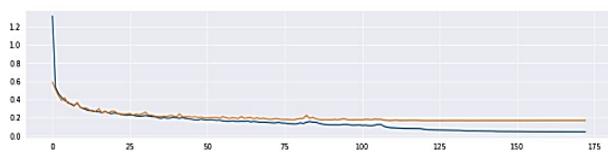

Fig. 13. Mean absolute error

$$EPOCH = 500$$
$$BATCH\_SIZE = 256$$
$$0.0419 - val\_loss: 0.1674 \qquad (5)$$

In this paper initially we also applied EDA model which was not able to derive as precise results than we have created a RNN based long short term memory (LSTM) algorithm with mean absolute error of 0.1878 which is able to predict. The less mean absolute error is there the more accurate model is, here our value is 0.18 which great for the model performance. It is derived out from the model that we have created a model which is able to regulate variables such as temperature, flow, pressure, speed, and other parameters.